\documentclass[fleqn,usenatbib]{mnras}

\usepackage{wrapfig,booktabs}

\usepackage{newtxtext,newtxmath}
\usepackage[T1]{fontenc}
\usepackage{ae,aecompl}

\usepackage{graphicx}	
\usepackage{amsmath}	

\title[ULXs in Edge-On Spiral Galaxies]{Ultraluminous X-ray Sources in Seven Edge-On Spiral Galaxies}

\author[K.C. Dage et al.]{
Kristen C. Dage,$^{1,2,3}$\thanks{E-mail: kristen.dage@mail.mcgill.ca}
Noah Vowell, $^{3,4}$
Erica Thygesen,$^{3}$
Arash Bahramian,$^{5}$
\newauthor
Daryl Haggard, $^{1,2}$
Konstantinos Kovlakas, $^{6}$
Arunav Kundu,$^{7}$
\newauthor
Thomas J. Maccarone,$^{8}$
Jay Strader,$^{3}$ 
Ryan Urquhart,$^{3}$ 
Stephen E. Zepf $^{3}$
\\
\\
$^{1}$ Department of Physics, McGill University, 3600 University Street, Montr\'eal, QC H3A 2T8, Canada\\
$^{2}$ McGill Space Institute, McGill University, 3550 University Street, Montr\'eal, QC H3A 2A7, Canada\\
$^{3}$Department  of  Physics  and  Astronomy,  Michigan  State  University,  East Lansing, MI 48824\\
$^{4}$Department of Natural Sciences, University of Michigan-Dearborn, 4901 Evergreen Rd. Dearborn, MI, 48128\\
$^{5}$ International Centre for Radio Astronomy Research $--$ Curtin University, GPO Box U1987, Perth, WA 6845, Australia\\
$^{6}$ 
Geneva Observatory, University of Geneva, Chemin des Maillettes 51, 1290
Versoix, Switzerland\\
$^{7}$Eureka Scientific, Inc., 2452 Delmer Street, Suite 100 Oakland, CA 94602, USA\\
$^{8}$ Department of Physics \& Astronomy, Box 41051, Science Building, Texas Tech University, Lubbock, TX 79409-1051, USA
}

\date{Accepted XXX. Received YYY; in original form ZZZ}

\pubyear{2020}

\begin{document}
\label{firstpage}
\pagerange{\pageref{firstpage}--\pageref{lastpage}}
\maketitle

\begin{abstract}
We investigate a sample of seven edge-on spiral galaxies using \textit{Chandra} observations. Edge-on spiral galaxies allow us to clearly separate source associated with their star-forming regions versus the outer edges of the system; offering a clear advantage over other systems. We uncover a number of X-ray point sources across these galaxies, and after eliminating contaminating foreground and background sources, we identify 12 candidate ultraluminous X-ray sources. All of these sources are projected onto the central regions, implying that the majority of ULXs in this sample of spiral galaxies are disk/bulge, and thus not halo sources. This also includes two transient ULXs, which may be long-duration transients and low mass X-ray binaries. This finding illustrates the need for further studies of transient ULXs.
\end{abstract}

\begin{keywords}
keyword1 -- keyword2 -- keyword3
\end{keywords}



\section{Introduction}

In this work, we seek to characterize the population of ultraluminous X-ray sources (ULXs) in and around spiral galaxies. ULXs are non-nuclear X-ray point source with X-ray luminosities exceeding $10^{39}$ erg s$^{-1}$ (the Eddington Limit for a 10 $M_{\odot}$ black hole assuming isotropic emission) \citep{1989ARA&A..27...87F}. Hundreds of ULX candidates have been identified (e.g.\citealt{2020MNRAS.498.4790K}, which finds 629 candidate ULXs in 309 galaxies within $D<40$ Mpc).  Most of the known ULXs have been identified in star-forming regions of galaxies (e.g. \citealt{2000MNRAS.315...98R, 2019ApJ...877...57Q} and references therein), although a small number have been identified in globular clusters associated with elliptical galaxies (see \citealt{2007Natur.445..183M,2019MNRAS.485.1694D} and references therein). 

A discovery that some of the ULXs associated with younger systems are found to show pulsations changed our understanding of ULXs, as the presence of pulsations rules out a black hole as the accretor for these sources, and points towards a neutron star (NS) as the primary accretor in the binary system (see e.g. \citealt{2014Natur.514..202B,2020MNRAS.495.2664C,  2021arXiv210311650Q}). 

Physical characteristics of ULXs can be studied via the spectral shape of their X-ray emission. Of the large numbers of ULXs studied and characterized, many of them fit into three classes: a broadened disk class, and hard ultraluminous and soft ultraluminous regimes. The typical power-law index of these sources ranges between 1.8-2.0, and the best fit model is an absorbed accretion disk plus power-law. Many sources show a soft excess in the lower energy regimes  \citep[e.g.][]{2006MNRAS.371..673G, 2009MNRAS.398.1450K, 2008ApJ...687..471B}, which are thought to be due to an optically thick outflow \citep{2003MNRAS.345..657K}.

These different spectral models are either mapped onto different accretion regimes determined by how greatly the source luminosity exceeds the Eddington limit (e.g. \citealt{2009MNRAS.397.1836G}) or different geometries relating to obscuration from the accreting material (see the recent review by \citealt{2017ARA&A..55..303K} and references therein). A rare subset of ULXs, ultraluminous supersoft sources (ULSs) have also been identified \cite{2016MNRAS.456.1859U}.

ULXs can have similar Eddington or super-Eddington luminosities, but show distinct spectral properties. ULSs typically show little-to-no emission above $\sim1$\,keV; they are dominated by a soft blackbody spectrum at $kT \sim 100$ eV \citep{2008ApJ...674L..73L}. In some ULSs, an additional weak power-law tail extending to higher energies is sometimes detected \citep[e.g.,][]{2011ApJ...737...87J}. The soft thermal emission in ULSs is often attributed to an optically-thick wind that obscures and reprocesses most hard emission that is emitted closer to the compact accretor \citep{2016MNRAS.456.1837S,2016MNRAS.456.1859U}; evidence of relativistic outflows have been detected in several ULSs \citep{2015Natur.528..108L, 2017MNRAS.468.2865P}. Some sources have even been observed switching between ULS and ULX modes, explained by changes in the line of sight absorption column, possibly a result of a clumpy wind or disk precession \citep{2017MNRAS.468.2865P}. Additionally, complete eclipses have been observed in two ULSs \citep{2002ApJ...574..382S,2016MNRAS.456.1859U} which also indicates these sources are being viewed close to edge-on. However, eclipses have also been observed in standard ULXs \citep{2016ApJ...831...56U}, suggesting that inclination angle cannot be the sole determination in whether a source appears as a ULS or ULX.

 \citealt{2016Natur.538..356I} discovered two sources, coincident with globular clusters, which flared in X-ray to above the Eddington limit, and then remained undetected. The physical cause behind these sources is not well characterized and difficult to study, but the nature of transient and flaring ULXs poses many interesting questions in astronomy \citep[e.g.][]{2018MNRAS.476.4272E, 2021MNRAS.501.1002W}.  While some have been identified and studied in depth (e.g. \citealt{2010ApJ...721..323S, 2012ApJ...747L..39H,2018ApJ...863..141B,  2020arXiv200100642E}, among others), including one discovered in the Milky Way \citep{2018Natur.562..233V}, it is unclear how these sources impact large scale studies of ULX populations.

 Of the ULXs associated with younger systems, most of these are in face-on spiral galaxies, where it is not possible to discern if the ULXs are associated with the disk, bulge, or halo population. By contrast, for edge-on spirals, the halo population
can be disentangled from the star forming disk, and the reddening issues which complicate studies of face-on spirals can be largely bypassed. Very few ULX systems in edge-on spirals have been studied in depth, among these include the pulsating ULX in NGC 5907 has been identified in an edge-on spiral galaxy \citep{2017Sci...355..817I}, and a source identified in NGC 891 by \citealt{2012ApJ...747L..39H}.

A survey by \cite{2009ApJ...703..159S} shows that ULXs are mainly associated with the star-forming regions of galaxies. Elliptical galaxies are known to host off-centre ULXs, but those are associated with globular clusters \cite{2021MNRAS.504.1545D}. Thus, the identification and detailed study of ULXs that are neither hosted by globular clusters, nor associated with the star-forming regions of the galaxies, provides a different angle on ULX systems.


In this body of work, we characterise ultraluminous X-ray sources in seven edge-on spiral galaxies. The paper is structured as follows: Section \ref{sec:obs} describes the spiral galaxy sample and data analysis, Section \ref{sec:res} describes the major results of this paper, and the impact of these results are discussed in Section \ref{sec:conclusions}.

\section{Observations and Analysis}
\label{sec:obs}
We utilize archival observations of seven edge-on spiral galaxies with inclination angles between 75 degrees and 90 degrees: NGC~891,  NGC~2683, NGC~3556, NGC~4013, NGC~4157, NGC~7331, and NGC~7814 (see Table \ref{galaxies} for a complete list of ObsIDs and galaxy distances used for luminosity calculations).

NGC~7331 has been observed many times with \textit{Chandra} (e.g. \citealt{2019Jin}), but this comparative study of many edge-on spirals focuses only on sources at the galaxy centre. 

NGC~891 and NGC~4013 were each observed twice with \textit{Chandra}. NGC~891 was observed in the year 2000 for 51ks and in the year 2011 for 2ks. The distance estimates were obtained using statistical distances from \cite{2021MNRAS.tmp.1552K}. A subset of these galaxies also have archival observations from \textit{XMM-Newton}, and event files from the Processing Pipeline System and can be utilised to track the ULX sources identified by \textit{Chandra}, and whether or not they are persistent. 

\begin{figure}
\includegraphics[width=8cm]{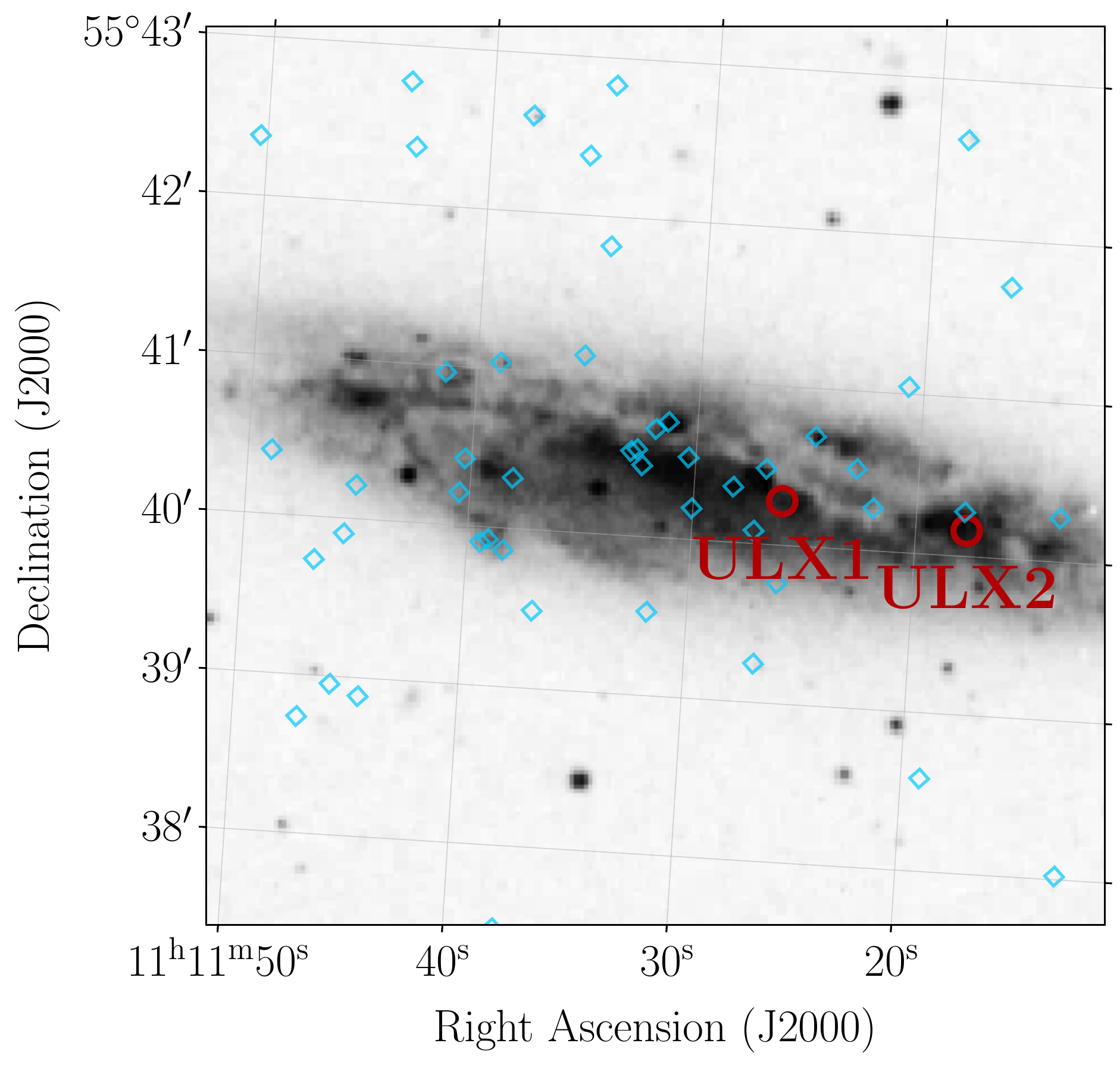}
\caption{DSS image of NGC~3556 with ULXs (open circles, 5" radius) and all other X-ray point sources identified (diamonds, 2.5" radius).}

\label{fig:im3556}
\end{figure}

\begin{figure}
\includegraphics[width=8cm]{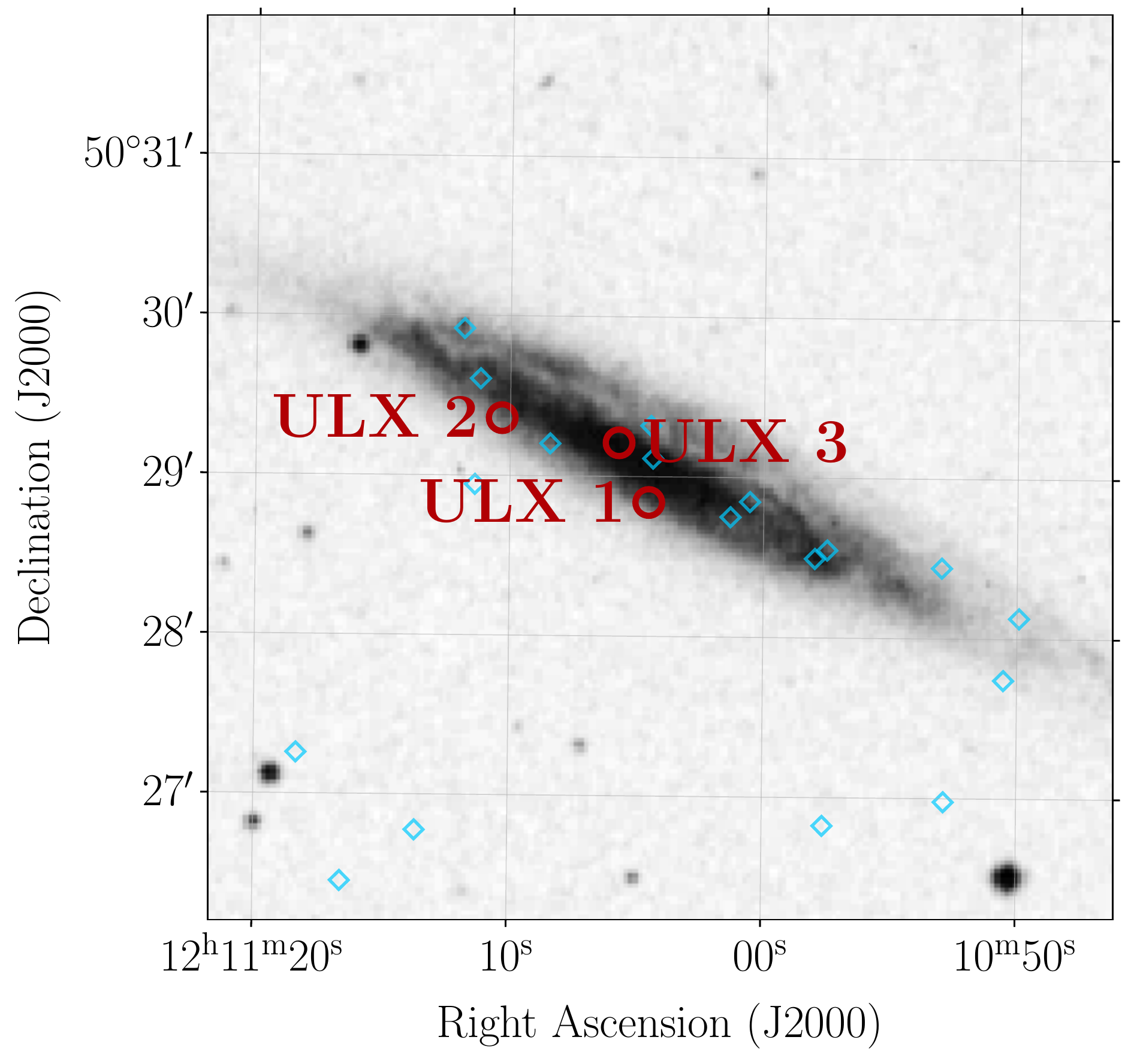}
\caption{ DSS image of NGC~4157 with ULXs (open circles, 5" radius) and all other X-ray point sources identified (diamonds, 2.5" radius).}

\label{fig:im4157}
\end{figure}

\begin{figure}
\includegraphics[width=8cm]{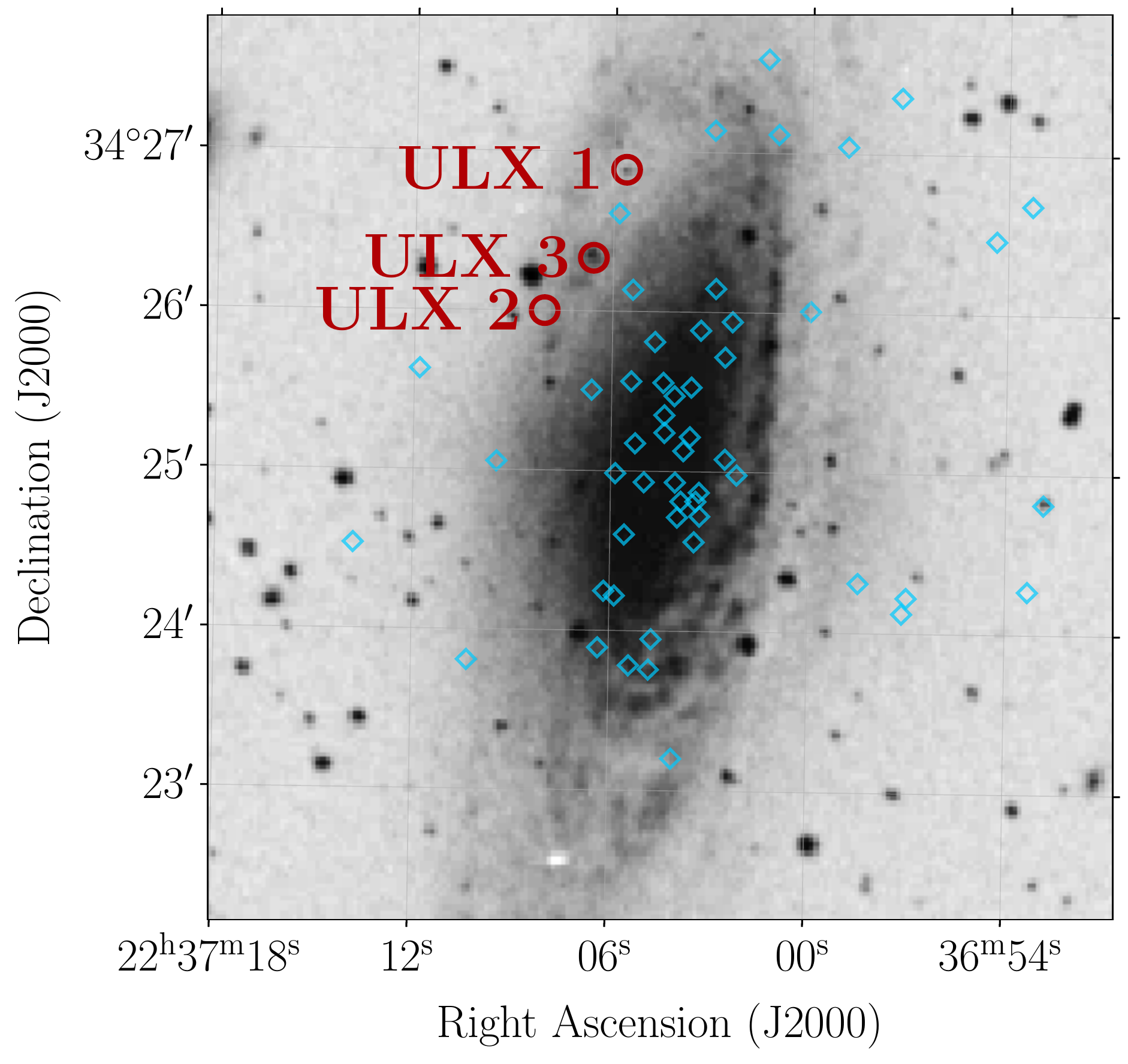}
\caption{DSS image of NGC~7331 with ULXs (open circles, 5" radius) and all other X-ray point sources identified (diamonds, 2.5" radius). See \citet{2007Abolmasov} for a discussion of the potential optical counterparts to the ULXs in the galaxy centre. }

\label{fig:im7331}
\end{figure}

\begin{figure}
\includegraphics[width=8cm]{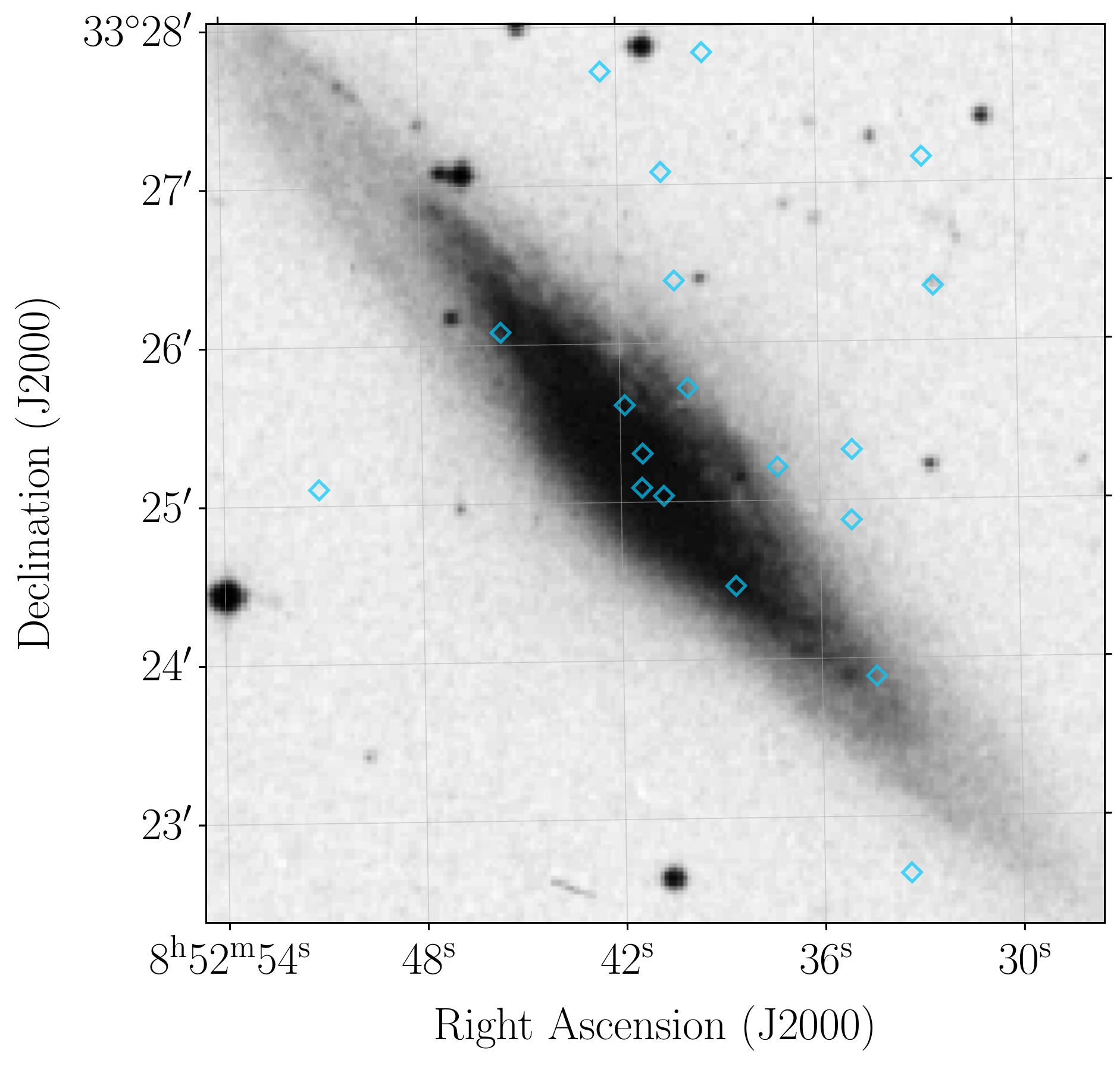}
\caption{DSS image of NGC~2683 with X-ray point sources identified (diamonds, 2.5" radius). No ULXs were identified in this galaxy.}

\label{fig:im2683}
\end{figure}

\begin{figure}
\includegraphics[width=8cm]{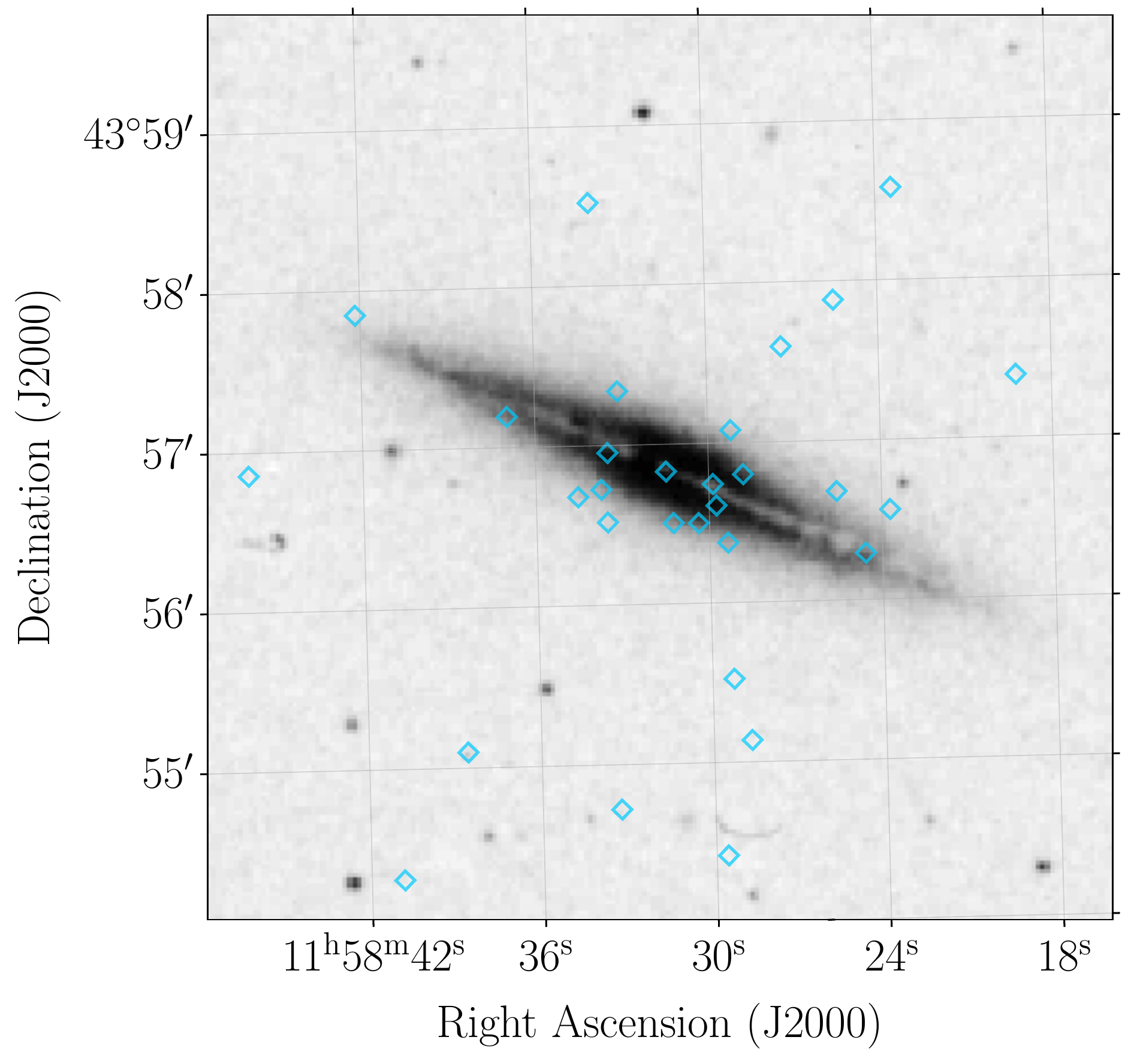}
\caption{DSS image of NGC~4013 with X-ray point sources identified (diamonds, 2.5" radius). No ULXs were identified in this galaxy.}

\label{fig:im4013}
\end{figure}

\begin{figure}
\includegraphics[width=8cm]{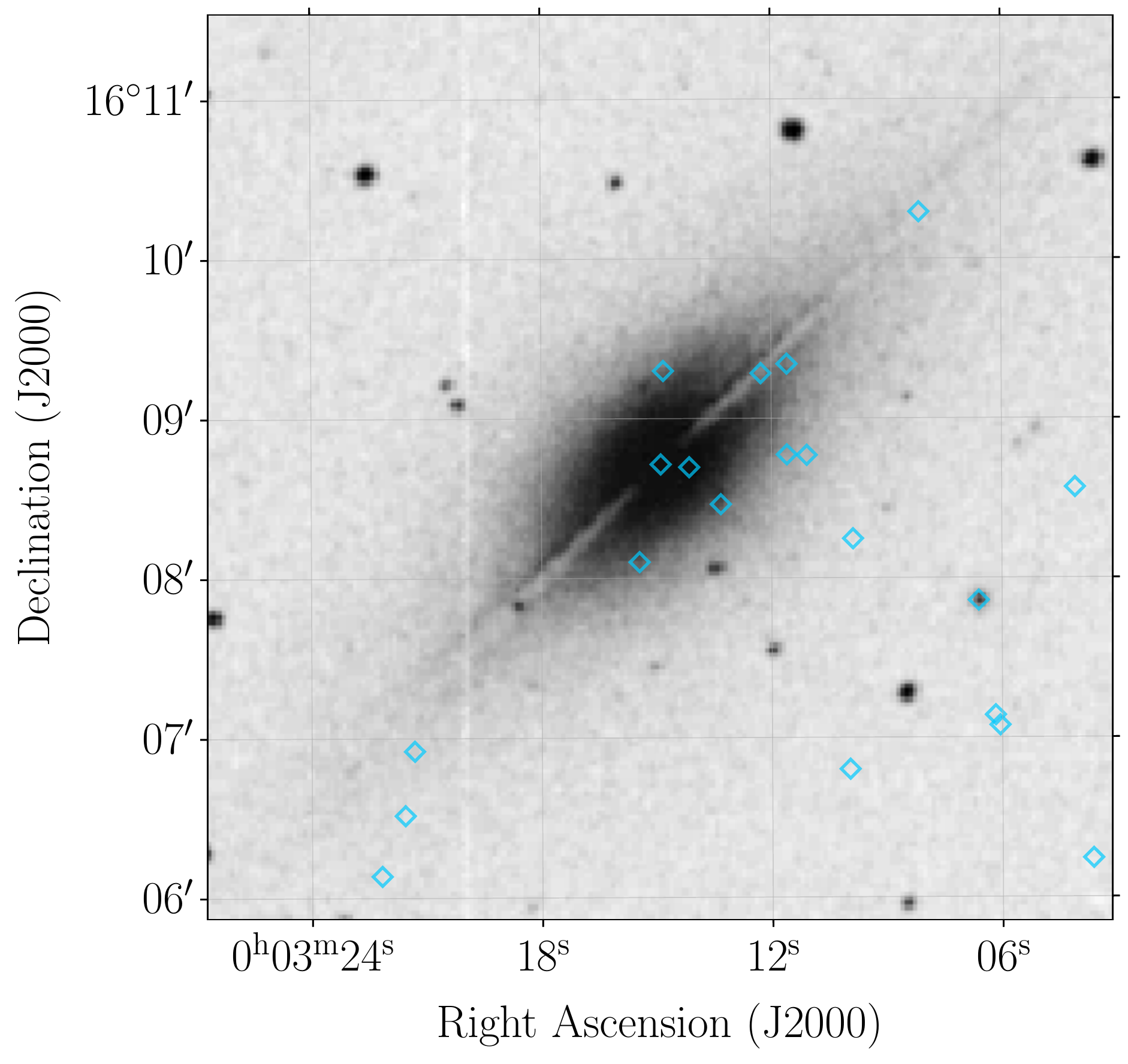}
\caption{DSS image of NGC~7814 with X-ray point sources identified (diamonds, 2.5" radius). No ULXs were identified in this galaxy.}

\label{fig:im7814}
\end{figure}

We use archival images from the ESO Online Digitized Sky Survey \footnote{\url{https://archive.eso.org/dss/dss}} to examine the location of the ULXs relative to the galaxy centre (see Figures \ref{fig:im891}-\ref{fig:im7814}). Note that analysis by \citealt{2007Abolmasov} suggests a stellar complex as an optical counterpart to one of to the ULXs in NGC 7331. 

\begin{table*}
\begin{centering}
\caption{Observations and distances of edge-on spiral galaxies in this paper. }
\label{galaxies}
\begin{tabular}{|l|l|l|l|l|l|l|l|}
\hline
\hline
Galaxy         & NGC~891 & NGC~2683 & NGC~3556 & NGC~4013 & NGC~4157 & NGC~7331 & NGC~7814 \\ \hline
ObsID          & 794, 14376   & 11311         & 2025          & 4013, 4739    & 11310         & 2198         & 11309         \\
Date & 2000-11-01,2011-12-20&2011-01-05 &2001-09-08  &2003-03-16,2004-02-03  &2010-08-21 & 2001-01-27&2009-09-01 \\
Observation Length (ksec) &51.0, 2.0 &38.5&59.4&4.9,79.1&59.3&29.5& 58.3\\ 
Distance (Mpc) & 9.11 $\pm$ 0.57        & 9.04 $\pm$ 0.86         &   9.90 $\pm$2.27& 19.71$\pm$ 3.46      & 16.95$\pm$3.07    & 14.40 $\pm$ 1.06     & 14.39 $\pm$ 1.99     \\ 
$n_H$ ($10^{20}$  cm$^{-2}$) &8.14&2.98&0.77&1.39&1.93&8.35&3.71\\
Diameter ($K_s$ Isophotal, arcsec) &225.4&153.2&184.6&111.4&134.9&155.8&102.3\\\hline

\end{tabular}
\end{centering}
\end{table*}
\subsection{Point Source Identification}

The X-ray point sources in this sample were identified using \textsc{ciao} \citep{2006SPIE.6270E..1VF}. We ran \texttt{wavdetect} on `broad' (0.5-7 keV) images to identify sources, with an PSF map centred at 2.3 keV. The wavelet scales were set to 1.0, 2.0, 4.0, 8.0 and 16.0 pixels. We used an encircled count fraction of 0.9 and a significance threshold of 10$^{-6}$, corresponding to about one false detection per chip.

We use \texttt{srcflux}\footnote{\url{http://cxc.harvard.edu/ciao/ahelp/srcflux.html}} to calculate fluxes for all of the sources detected in the observations. The background regions were automatically created by \texttt{roi}\footnote{\url{http://cxc.harvard.edu/ciao/ahelp/roi.html}}. We froze the equivalent hydrogen absorption column to the value for the line of sight of the Galaxy ($n_H$) \footnote{\url{http://cxc.harvard.edu/toolkit/colden.jsp}} (See Table \ref{galaxies}), and estimated the flux with a fixed power-law index of 1.7 to all the sources. Given that this method can slightly underestimate the flux (and does not take uncertainties into account), we selected all of the sources with an estimated $L_X$ of ~7$\times 10^{38}$ erg/s or above. We searched existing catalogs (including  \citealt{2005ApJS..156...35E, 2006AJ....131.1163S, 2009yCat.2294....0A,2010AA...518A..10V,2018yCat.1345....0G}) to remove contaminants such as QSOs, AGN and foreground stars. The remaining sources were fit with \textsc{xspec} to measure the flux of the sources more accurately, and to determine their spectral properties, and the 12 ULX candidates are listed in Table \ref{ulxcandidates}, along with bright contaminants.








\begin{table*}
\centering
\caption{12 ULX candidates identified from point source catalogs. Many of these sources have been identified in previous surveys, and several sources have been studied in depth, including NGC 891 ULX1 (\citealt{2012ApJ...747L..39H, 2018ApJ...866..126H}), and the NGC 7331 ULXs \citep{2019Jin}, but for the purposes of this study we fit all the sources with the same spectral models for cross-comparison.}
\label{ulxcandidates}

\begin{tabular}{l||l|l|l|l|l|l|}
\hline
\hline
Galaxy &RA         & Dec         & Reference                  & Projected on Galaxy? &ObsID & Name  \\ \hline
NGC 891 &2:22:33.45$\pm$0.50 & 42:20:26.88$\pm$0.50 & \citet{2012ApJ...747L..39H} & Yes & 14376 only  & N891ULX1\\ 
NGC 891 &2:22:33.92$\pm$0.53& 42:20:53.42$\pm$0.40 & - & Yes & 14376  only & N891ULX2 \\ 
NGC 891 &2:22:31.31$\pm$0.69& 42:19:57.39$\pm$0.55& \citet{2004ApJS..154..519S}  & Yes   &794, 14376 (faint)  & N891ULX3 \\ 
NGC 891 &2:22:31.41$\pm$0.65& 42:20:23.97$\pm$0.55 & \citet{2006ApJS..166..154P}  & Yes       &794, 14376 (faint)  & N891ULX4       \\ 
NGC 891 &2:22:24.42$\pm$0.55& 42:21:38.43$\pm$0.5& \citet{2010AA...518A..10V} &  AGN    &794        & -     \\ 
NGC 891 &2:22:25.31$\pm$1.74& 42:24:50.78$\pm$1.05 & \citet{2005ApJS..156...35E}  & QSO &794         &-     \\ \hline
NGC 2683 & 8:52:41.31$\pm$1.56& 33:25:18.42$\pm$1.11 & \citet{2006AJ....131.1163S}&  AGN & 11311 & - \\\hline
NGC 3556 & 11:11:26.03$\pm$0.96 &55:40:16.66$\pm$0.55& \citet{2011ApJ...741...49S}& Yes   & 2025 & N3556ULX1\\
NGC 3556 & 11:11:17.72$\pm$1.46 & 55:40:09.79$\pm$0.77 & \citet{2003ApJ...598..969W}& Yes & 2025 & N3556ULX2 \\\hline
NGC 4013 & 11:58:45.44$\pm$1.49 & 43:59:59.97$\pm$0.88 & \citet{2018yCat.1345....0G}&  AGN & 4739 & - \\
NGC 4013 &11:58:38.56$\pm$1.91 & 43:55:05.85$\pm$1.05 & \citet{2018yCat.1345....0G}& QSO& 4739 &- \\
NGC 4013 & 11:58:28.68$\pm$1.38 & 43:55:08.15$\pm$0.73 & \citet{2009yCat.2294....0A}&  AGN & 4739 &- \\\hline
NGC 4157 &12:11:04.53$\pm$1.49 &50:28:50.33+$\pm$.82 &-  & Yes & 11310&N4157ULX1 \\
NGC 4157 & 12:11:10.33$\pm$1.73&50:29:21.67$\pm$1.12 & -& Yes & 11310&N4157ULX2 \\ 
NGC 4157 & 12:11:05.72$\pm$1.84&50:29:12.65$\pm$1.25 & -& Yes & 11310 &N4157ULX3 \\ 

NGC 4157 &  12:11:18.21$\pm$1.08& 50:26:52.35$\pm$0.70 &\citet{2007AJ....134.1403R}&Star?  & 11310 &-\\ 
NGC 4157 & 12:11:13.65$\pm$0.70 &50:26:46.62$\pm$0.57 & \citet{2018yCat.1345....0G} & QSO & 11310 &-\\
NGC 4157 &12:11:16.96 +1.25 &50:22:05.54+0.94 &\citet{2018yCat.1345....0G} & Star & 11310 &-\\\hline

NGC 7331 & 22:37:05.63$\pm$0.81 & 34:26:53.20$\pm$0.62 & \citet{2004ApJS..154..519S} & Yes& 2198  &N7331ULX1\\

NGC 7331 & 22:37:08.07$\pm$0.64 & 34:26:00.00$\pm$0.55 &\citet{2004ApJS..154..519S}  &Yes  & 2198&N7331ULX2\\
NGC 7331 &22:37:06.60$\pm$0.93 &34:26:19.94$\pm$0.34 &\citet{2004ApJS..154..519S}  & Yes & 2198&N7331ULX3 \\
NGC 7331 &22:36:51.78$\pm$1.29 &34:26:00.39$\pm$0.59 & \cite{2007AJ....134.1403R}& Star? & 2198 &-\\\hline
NGC 7814 &0:03:13.31$\pm$1.15 & 16:08:27.71$\pm$0.65& \citet{2018yCat.1345....0G}&Star& 11309 &- \\
NGC 7814 & 0.02:49.45$\pm$1.53 & 16:10:25.83$\pm$1.04 & \citet{2018yCat.1345....0G}& Star & 11309 & -\\ \hline
\end{tabular}
\end{table*}

\subsection{Spectral Fitting of ULX Candidates}
A number of the sources identified in the previous section were found to have X-ray luminosities exceeding the Eddington Limit  for a 10 $M_\odot$ black hole ($\sim 10^{39}$erg/s). Their spectra were extracted with \textsc{ciao}, grouped by bins of 20, and fit with in detail \textsc{xspec}. Sources with fewer than 100 counts were binned by 1 and fit using W-statistics \citep{1979ApJ...228..939C}. The sources were fit either with \textsc{tbabs*tbabs*pegpwrlw}, an absorbed power-law model or \textsc{tbabs*tbabs*diskbb}, an absorbed multi-black body component model \citep{1984PASJ...36..741M, 2000ApJ...542..914W}. The best fit values are presented in Table \ref{ulxfits}. The line of sight absorption column was fixed with line-of-sight Galactic values (see Table \ref{galaxies}) and abundance of elements from \citealt{2000ApJ...542..914W}. In any cases where the secondary absorption was consistent with zero, it was removed from the model. The best fit values for the secondary absorption column where fit are presented in footnotes by the best fit value for the spectral parameter ($\Gamma$ or $T_{in}$).

\begin{table*}
\centering
    \caption{\textit{Chandra} Fit Parameters and Fluxes (0.5-8 keV) for spectral best fit single-component models, \texttt{tbabs*diskbb} and \texttt{tbabs*pegpwrlw} for ULX candidates. Hydrogen column density ($N_H$) frozen to Galactic values (see Table \ref{galaxies}). Lower count observations fit with C-stat have their statistics presented in parentheses. All fluxes shown are unabsorbed. Sources marked with $\dagger$ have significant $n_H$, best fit values are presented in Table \ref{nhtable}.  NGC 4157 ULX4 (marked with $^o$, was too off-axis to be fit with \textsc{xspec}, and instead we use \textsc{pimms} to estimate the flux based off of the background-subtracted count rate, 1.9 $\times 10^{-3}$ $\pm$ 1.8$\times 10^{-4}$ counts/sec, assuming a fixed power-law index of $\Gamma$=1.7. }
\label{ulxfits}
\begin{tabular}{|l|clc|lclc|}
\hline
\hline
\toprule
  &    & & NGC~891\ \\ \hline
  &       \multicolumn{3}{c}{\textsc{tbabs*pegpwrlw}} & \multicolumn{4}{|c}{\textsc{tbabs*diskbb}} \\ \hline
\midrule
 Source(ObsID)  & $\Gamma$ &   $\chi^2_{\nu}$/d.o.f. & PL Flux & $T_{in}$ &  Disk Norm &  $\chi^2_{\nu}$/d.o.f. &  Disk Flux  \\
&& \textbf{or} (C-stat)&($10^{-13}$ $\frac{\mathrm{erg}}{ \mathrm{cm}^{2} s}$)& (keV)&($10^{-2}$)& \textbf{or} (C-stat)&($10^{-13}$ $\frac{\mathrm{erg}}{ \mathrm{cm}^{2} s}$)\\ \hline  \hline

N891ULX1(14376) & 1.5 ($\pm 0.4$) & 1.71/4  & 8.8 ($\pm 0.2$)  & 1.3 ($^{+0.7}_{-0.4}$)  & $\leq$ 11.72  & 1.68/4 & 6.7 ($\pm 2.0$)  \\ 
N891ULX2(14376) & 1.5 ($\pm0.4$) & (29.80/50) & 3.5 ($^{+1.2}_{-0.9}$) & 1.4 ($^{+0.9}_{-0.4}$)& $\leq$ 4.3& (30.15/50) &3.0 $\pm 1.1$ \\

N891ULX3(794) $\dagger$ & 2.2 ($\pm0.2$)&1.64/68 & 4.6 ($\pm 0.4$) &1.22 ($\pm 0.1$)&  6.9 $\pm 0.4$&2.16/68  & 3.0($\pm 0.2$)\\
N891ULX4(794) $\dagger$ & 1.9 ($\pm0.1$)  & 1.02/86& 7.0 ($\pm 0.5$) & 1.6($\pm 0.1$) &0.5 ($\pm 0.2$) & 1.22/86 & 5.2 ($\pm0.3$)\\

\hline
\toprule 
 &&& NGC~3556\\ \hline
\midrule
N3556ULX1(2025)  $\dagger$ & 1.8($\pm0.2$)  &1.04/53  & 2.5 ($\pm0.2$) &  1.7 ($\pm 0.2$) & 1.32/54  & 1.3 ($\pm 0.3$)& 1.8 ($\pm 0.14$)  \\ 
N3556ULX2(2025) &  1.5 ($\pm 0.2$)& 0.71/10 & 0.4 ($\pm 0.1$)   &0.9 ($^{+0.3}_{-0.2}$)& 1.5 ($\pm 0.2$) & 0.82/10& 0.23 $\pm0.06$ \\
\hline

\toprule 
 &&& NGC~4157\\ \hline
\midrule
N4157ULX1(11310)$\dagger$  &1.8 ($\pm 0.2$) &0.95/52  &3.5 ($\pm 0.3$)  &1.6 ($\pm 0.2$)  &2.3 ($^{+0.1}_{-0.9}$)  & 1.10/52& 2.6 ($\pm0.2$)  \\ 
N4157ULX2(11310)   &1.1 ($\pm 0.4$) &1.87/4 & 0.34 $\pm 0.08$  & - &-  & -& - \\
N4157ULX3(11310) $^o$  &1.7 (fixed) & -  & 0.15 $\pm$ 0.12 &-& -&-&- \\
\hline

\toprule 
 &&& NGC~7331\\ \hline
\midrule
N7331ULX1(2198) & 1.4 ($\pm 0.3$) & 1.69/3 & 0.45 ($\pm 0.09$)  &1.5 ($^{+0.8}_{-0.4}$)& $\leq$ 0.4 &0.90/43& 0.37 $(\pm 0.09)$ \\

N7331ULX2(2198) &1.8 ($\pm0.3$) &0.86/6 &0.58 ($\pm 0.09$)  & 1.0 ($\pm 0.3$)&2.4 ($^{+0.3}_{-0.1}$)&1.28/6 & 0.47 ($\pm 0.08$) \\
N7331ULX3(2198)  & 1.4 ($\pm 0.4$)& 0.55/7 & 0.75 ($\pm0.01$)  &1.4 ($^{+0.9}_{-0.4}$)& $\leq$ 0.8& 0.64/5 & 0.6 ($\pm0.2$) \\ \hline

\end{tabular}
\end{table*}

\begin{table}
\caption{Best fitting secondary absorption column density for ULX candidates in units of $10^{22}$ cm$^2$, when intrinsic absorption is not negligible.}
\label{nhtable}
\begin{tabular}{|l|l|l|}
\hline
Source        & Power-Law $n_H$   & Disk $n_H$        \\ \hline
NGC 891 ULX3  & 0.47($\pm 0.08$)    & 0.19 ($\pm 0.05$) \\ \hline
NGC 891 ULX4  & 0.64 ($\pm 0.08$) & 0.37 ($\pm 0.06$) \\ \hline
NGC 3556 ULX1 & 0.43 ($\pm0.09$)  & -                 \\ \hline
NGC4157 ULX1  & 0.81 ($\pm0.16$)  & 0.27 ($\pm0.08$)  \\ \hline
\end{tabular}
\end{table}

\begin{figure}
\includegraphics[width=8cm]{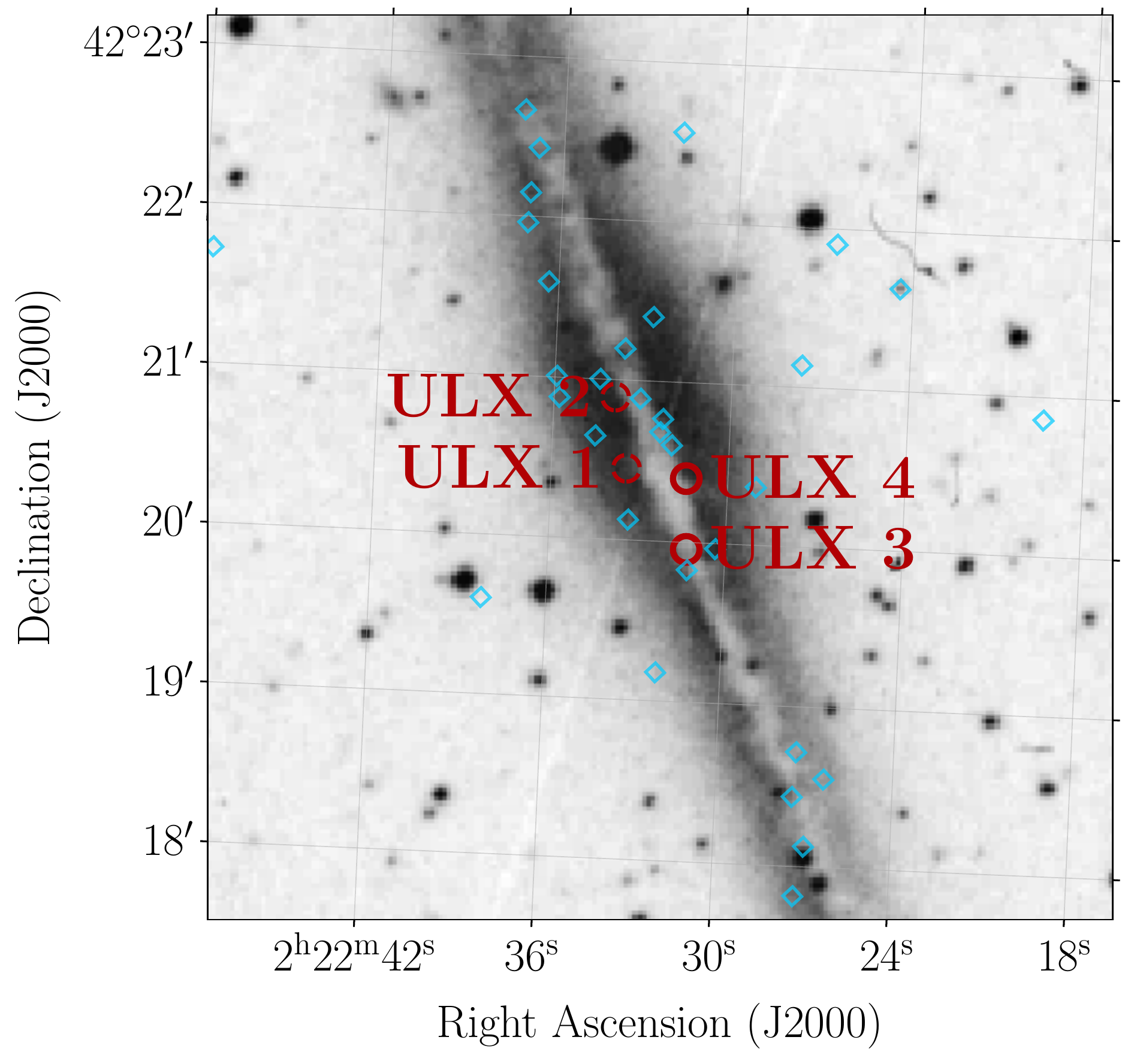}
\caption{ DSS image of NGC~891 with ULXs (open circles, 5" radius) and all other X-ray point sources identified (diamonds, 2.5" radius). ULX 1 and ULX 2, marked with dashed lines, are both transient sources (see Section 2.4).  }

\label{fig:im891}
\end{figure}


\subsection{NuSTAR Observations of NGC 4157}
NGC 4157 was observed by the Nuclear Spectroscopic Telescope Array (\textit{NuSTAR}) \citealt{2013ApJ...770..103H} for $\sim$ 40ks on 2020-10-19 (program number 06058, ObsIDs 30601015001, 30601015002). 
The NuSTAR data was reduced using \textsc{nustardas} version 2.0.0 \citep{2015ApJS..220....8M}, included in \textsc{HEASOFT} 6.28. We extracted source and background spectra in the 3-78 keV band following standard procedures outlined by the NuSTAR guidelines \footnote{NuSTAR data analysis software users guide, v1.9.6 \url{https://heasarc.gsfc.nasa.gov/docs/nustar/analysis/}}, using 60'' radius circular regions for both source and background from each detector. The spectra were binned to contain at least 20 source counts per bin and were fit jointly in \textsc{xspec}. Although ULX 1, the host galaxy, and nearby X-ray sources are blended together, the integrated spectrum contains useful information on the high-energy behaviour of this system, and due to NuSTAR's timing capabilities, we are able to place constraints on the presence of pulsations in the region of interest. The data were calibrated and barycentre corrected using {\it NuSTAR}/FPM calibration database version 20210315. 
We extracted source and background light curves and performed a standard pulsation search in the 3-79 keV band using \textsc{MaLTPyNT} \citep{2015ascl.soft02021B}, searching for possible pulsations, seen in some ULXs in other Galaxies \citep[e.g.][]{2014Natur.514..202B}. We find no clear evidence of pulsations in the range of $10^{-3}$ and 100 Hz and estimate a 5-sigma upper limit of 20\% RMS \citep{1989ARA&A..27..517V}. We note that this result is based on a standard search, and not an acceleration search.We fit the spectrum jointly with the FPMA and FPMB detectors and a power-law model \ ($n_H$ was frozen to NGC 4157's line-of-sight abundances) returned a best-fit power-law index of 1.96 $\pm$ 0.2, which is consistent with the best-fit parameter from the \textit{Chandra} observation, suggesting that ULX 1 may be dominating the flux.

\subsection{Follow Up on Transient Sources}

The number of ULXs in a given galaxy appears to be correlated the star-formation rate of the galaxy \citep[and references therein]{2009ApJ...703..159S,2012MNRAS.419.2095M}, as well as the metallicity \citep{2013ApJ...769...92P}. However, the role that transient ULXs play in this relationship is unclear, as is the cause of persistent versus transient sources. 

We utilized archival \textit{XMM-Newton} data to determine if these sources were observed in previous or subsequent archival observations in the data archive. Because \textit{XMM-Newton} does have a much lower spatial resolution than \textit{Chandra} blending of nearby sources is a major concern, and we do not extract and fit spectra. We do not consider new ULXs that may appear in \textit{XMM-Newton} data, as there is no good way to quantify the confusion effect. Instead, we only consider cases where the count rate of at the position of the previously identified ULX sources is a non-detection, or consistent with the background. If a source appears to have a clear X-ray point source associated with it in the \textit{XMM-Newton} image, then we consider it to be a persistent source over the time range probed in this study and do not conduct further analysis.

Currently archival data covers approximately a 17 year span of NGC~891 and a 6 year span of NGC~4157. NGC~891 was observed by XMM-Newton on 2002-08-22 (ObsID 0112280101), 2011-08-25 (ObsID 0670950101), and observed five times with in a month in 2017 (ObsIDs 078076101-078760501), as well as twice by Chandra on 2000-11-01 (ObsID 794) and on 2011-12-20 (ObsID 14376). NGC~4157 was observed by XMM-Newton on 2004-05-17 (ObsID 0203170101) and by Chandra on 2010-08-21 (ObsID 11310). The ULXs in NGC~4157 were detected in all of the the Chandra and XMM-Newton observations, but only two of the ULXs in NGC~891 were persistent across the archival data.

One of the NGC 891 sources, identified in \citealt{2012ApJ...747L..39H} was not observed in 2000 or 2002, but had reached a high X-ray luminosity in the 2011 data and remained bright in the 2017 observations \citep{2018ApJ...866..126H}. 

\begin{figure}
\includegraphics[width=8cm]{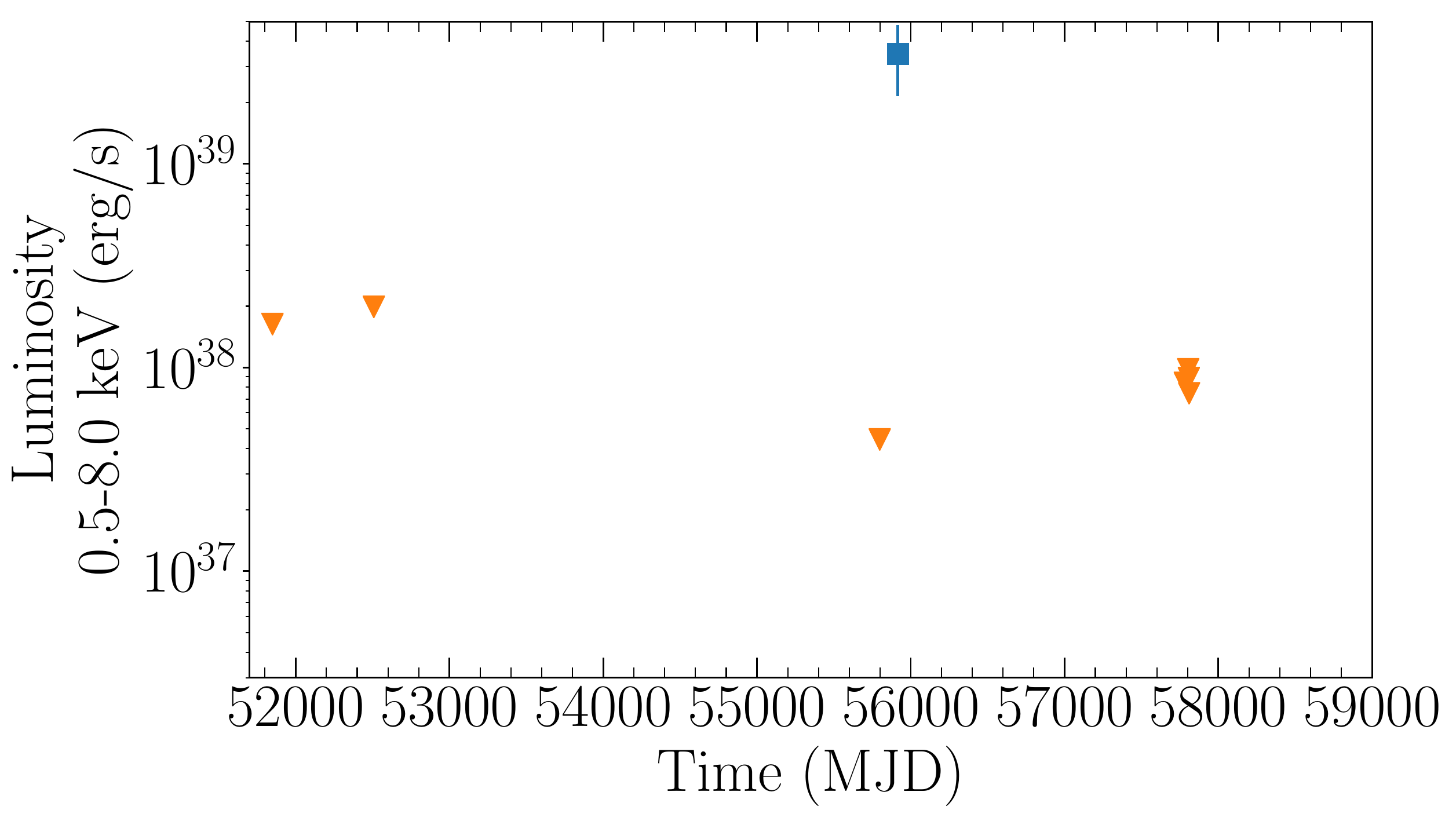}
\caption{ $L_X$ versus observation date (MJD) for NGC~891 ULX2. The blue square is the Chandra observation, and the orange triangles represent upper limits from the XMM-Newton observations.}

\label{fig:flare}
\end{figure}

The other transient source, NGC~891 ULX2 was only detected in the second of the Chandra observations (ObsID 14376 taken in December, 2011), but was not detected in all of the previous and subsequent data sets (see Figure \ref{fig:flare}). 

  We estimated the upper limits on the non-detections in NGC 891 ULX2 in the following manner: for Chandra ObsID 794, the upper-limit calculation is based on 3$\sigma$ of 100 counts (from a measured background of 71 counts), using statistics from \cite{1986ApJ...303..336G}. The count rates were converted to a flux using \textsc{pimms}, assuming a power-law index of 1.7 and $n_H$ consistent with the line-of-sight to the galaxy. The XMM-Newton upper limits were calculated using FLIX\footnote{\url{https://www.ledas.ac.uk/flix/flix.html}} (with the exception of observation 078076101, for which FLIX gave no limit), and converted to the 0.5-8 keV range using \textsc{pimms} and a power-law index of 1.7. 
 We estimate that NGC 891 ULX2 is consistently detected below 2$\times10^{38}$ erg/s, see Table \ref{table:xmm}. 
Based on the data in hand, the upper limit for the duration of the bright phase of NGC~891 ULX2 is at the very most approximately 5 years. The background subtracted lightcurve of the source was extracted using \textsc{ciao's dmextract}, and binned by 1, 10, 50 and 100 seconds but revealed no evidence of short-term variability.

\begin{table*}
\caption{Upper limits on non-detections of NGC~891 ULX2 in archival data. All observations are from \textit{XMM-Newton} unless indicated otherwise. }
\label{table:xmm}
\begin{tabular}{|l|l|l|l|l|l|l|l|}

\hline
ObsID         & Date       & ObsLen (ks) &  Flux (erg/s/cm$^2$)    & $L_X$ (erg/s)\\ \hline
794 (Chandra) & 2000-11-01 & 57.0   & $\leq$ 1.4e-14 & $\leq$1.4 $\times 10^{38}$\\ 
0112280101    & 2002-08-22 & 19.0         & $\leq$1.7e-14 & $\leq$1.6$\times 10^{38}$\\ 
0670950101    & 2011-08-27 & 133.0                &$\leq$ 3.8e-15 & $\leq$3.7$\times 10^{37}$\\ 
078076201     & 2017-01-29 & 73.0       & $\leq$7.2e-15 &$\leq$ 7.1$\times 10^{37}$\\ 
078076301     & 2017-02-24 & 74.0             & $\leq$7.6e-15 & $\leq$7.5$\times 10^{37}$\\
078076401     & 2017-02-20 & 74.4          & $\leq$8.4e-15 & $\leq$8.3$\times 10^{37}$\\
078076501     & 2017-02-26 & 73.8       & $\leq$6.4e-15 & $\leq$6.4$\times 10^{37}$\\\hline
\end{tabular}
\end{table*}

\subsection{Constraints on Optical Counterparts}
Regions of these galaxies have been imaged by various instruments on the HST using a variety of narrow and wide band filters. Where possible we accurately aligned the HST and Chandra astrometry using the positions of all the X-ray point sources detected in the program galaxies (assuming positional uncertainties returned by \texttt{wavdetect}) and inspected the optical image at the location of the candidate ULXs in order to identify optical counterparts. Details of the optical data will be presented in a future study of all X-ray binaries in these galaxies.

NGC891ULX1,  NGC891ULX3, and NGC891ULX4 lie within the field of view of deep 7712s ACS/WFC exposures in the F606W and F814W bands obtained by HST program GO 91414. NGC891ULX1 and NGC891ULX3 are coincident with point like optical sources in or near dust lanes. NGC891ULX3 appears to be significantly bluer than other point sources in the immediate neighbourhood. NGC891ULX4 does not have a resolved counterpart. NGC891ULX2 is located within the F450W and F814W WFPC2 images obtained by HST program SNAP 9042 and does not have an optical counterpart.

The NGC 3556 ULX candidates are located within a short exposure (160s) F606W WFPC2 images obtained by HST program SNAP 5446. Both of the candidates are coincident with optical candidates located at the outer edge of the Chandra positional accuracy. Due to the short exposure combined with the smaller field of view of the WFPC2 we could not refine the astrometric alignment of the HST and Chandra images using other X-ray sources in the field. The possible optical counterpart of NGC3556ULX1 is extremely bright M$_V$ $\approx$ -11 mag if it is located in NGC 3556 and not a foreground star projected on the field. Such a luminous object would be among the handful of most massive globular clusters observed in any galaxy, well into the regime of 'ultra compact dwarfs' that are tidally stripped nuclei.

There are no HST observations of the NGC 4157 ULX candidates. As reported previously by \cite{2007Abolmasov} NGC7331ULX3 is associated with a stellar complex in a deep F814W ACS image obtained by HST program GO 15645. NGC7331ULX2 does not appear to have an optical counterpart. The location of NGC7331ULX1 has been imaged in the narrow band F658N filter by WFPC2 as part of HST program GO 11966. It does not have an optical counterpart.

\section{Results}
\label{sec:res}

In this work, we present a detailed spectroscopy of new and previously discovered ULX sources identified in \textit{Chandra} observations of seven edge-on spiral galaxies. For sources with luminosity estimates exceeding the Eddington limit for a 10 $M_{\odot}$ black hole ($10^{39}$ erg/s), we match existing catalogs to determine if they have previously been identified as AGN or QSO. 

We compare the ULX candidates to DSS images. The majority of the ULX candidates were aligned with the disk/star-forming region of the galaxy, with four having no clear optical counterpart. Because of the lack of optical counterparts, it is unclear if these sources are X-ray binaries, background galaxies or foreground stars.  

\begin{figure}
\includegraphics[width=8cm]{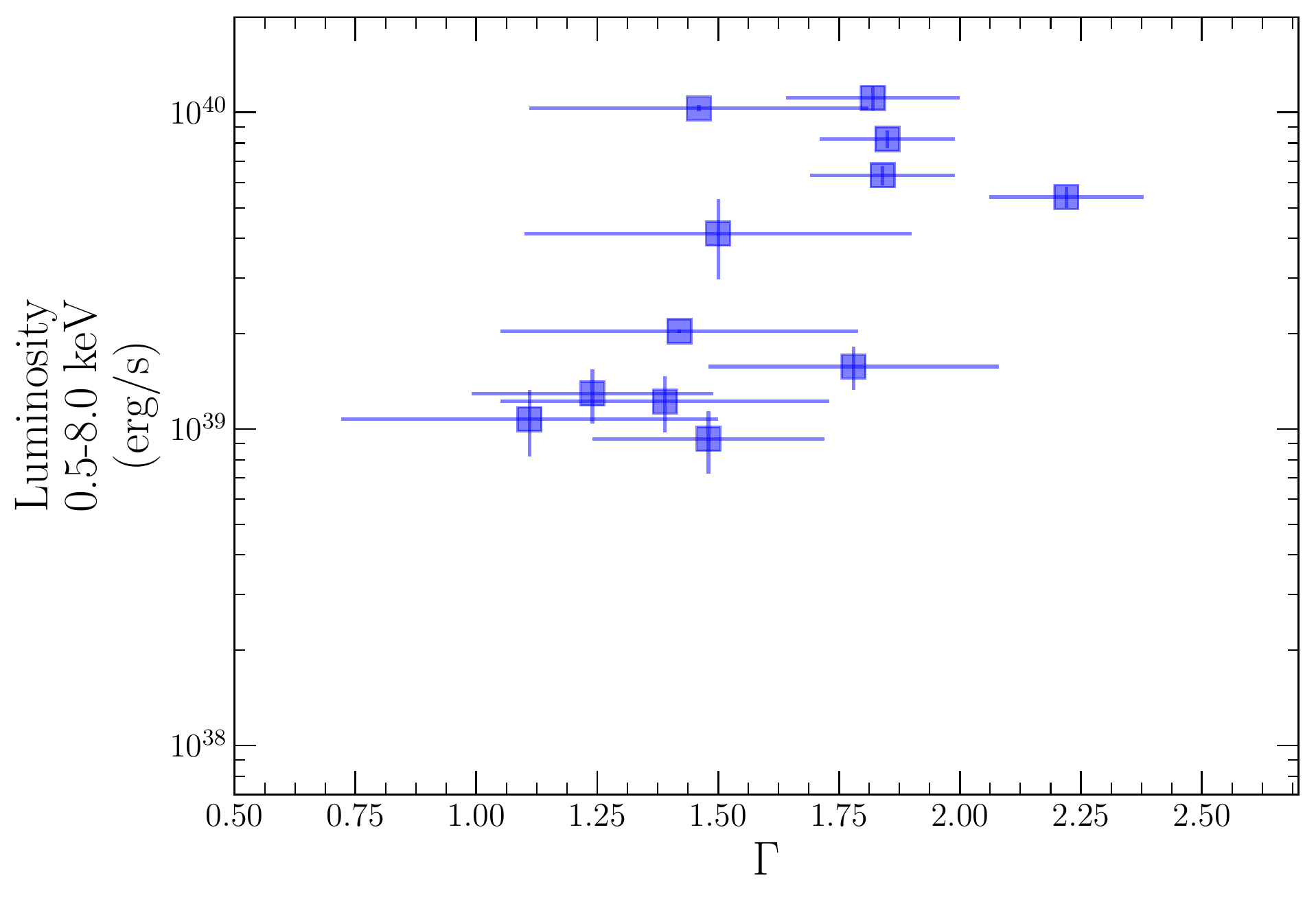}
\caption{ $L_X$ versus $\Gamma$ for ULXs best fit by a power-law model. See Table \ref{ulxfits} for measurements.}

\label{fig:lxg}
\end{figure}

\begin{figure}
\includegraphics[width=8cm]{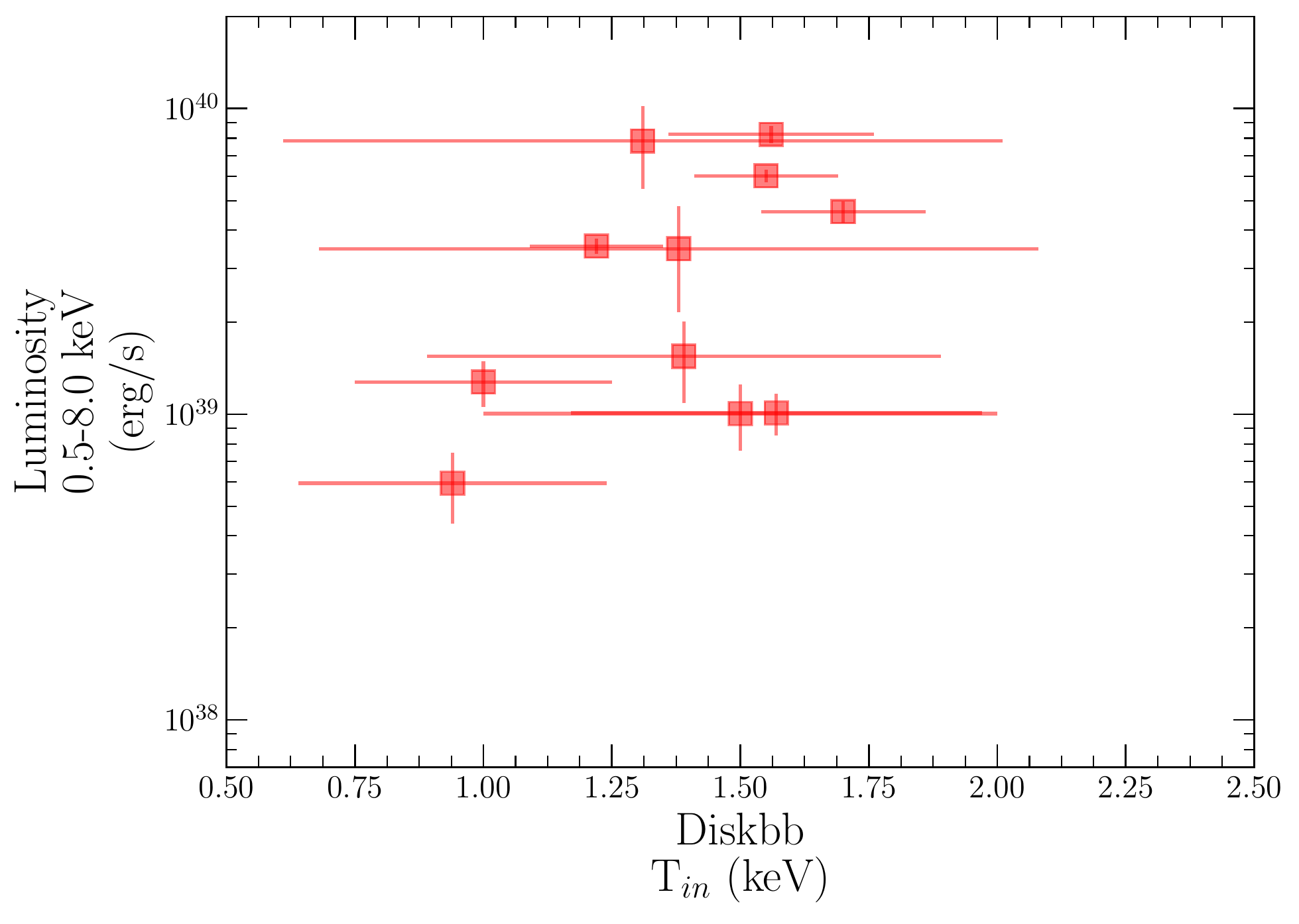}
\caption{ $L_X$ versus $T_{in}$ for ULXs best fit by a blackbody disk. See Table \ref{ulxfits} for measurements. }

\label{fig:lxtin}
\end{figure}

We perform detailed X-ray spectroscopy on the ULX candidates to determine the nature of their X-ray emission, and fit them with absorbed power-law and absorbed blackbody disk models. The majority of the sources were equally well fit by either model, however, a handful of sources were harder and not well-fit by the disk model. This could be due to the high level of galactic absorption making it difficult to accurately model the soft components of the emission.  The sources fit a range of X-ray luminosities and spectral parameters. 
The best fitting spectral parameters and X-ray luminosity are presented in Figures \ref{fig:lxg} and \ref{fig:lxtin}.
\begin{table}
\caption{$L_X$ of sources, calculated from the measured fluxes in Table \ref{ulxfits} assuming distances from Table \ref{galaxies}. }
\label{lxtable}
\begin{tabular}{lll}
\hline
Source         & Power-Law $L_X$                 & Disk $L_X$        \\
& $\times 10^{39}$ erg s$^{-1}$ &$\times 10^{39}$ erg s$^{-1}$ \\ \hline
NGC 891 ULX1   & 10.30 ($\pm$ 0.23)              & 7.83 ($\pm$ 2.37) \\
NGC 891 ULX2   & 4.14 ($\pm$ 1.17) & 3.47 ($\pm$ 1.32) \\
NGC 891 ULX3   & 5.41 ($\pm$0.41)                & 3.55 ($\pm$0.20)  \\
NGC 891 ULX4   & 8.24 ($\pm$ 0.53)               & 6.03 ($\pm$ 0.29) \\ \hline
NGC 3556 ULX1   &   6.34 ($\pm$0.44)       &    4.60 ($\pm 0.36$)               \\
NGC 3556 ULX2  &         0.93 ($\pm$0.21)      &          0.59 ($\pm 0.16$)        \\ \hline
NGC 4157 ULX1   &  11.11 ($\pm 0.95$)                     &       8.24 ($\pm 0.54$)         \\
NGC 4157 ULX2  &                1.07 ($\pm 0.25$)      & \\   
NGC 4157 ULX3  &      1.99 ($\pm 1.51$)                  &             -      \\ \hline
NGC 7731 ULX1  &                1.22 ($\pm 0.24$)       & 1.01 ($\pm 0.24$)                 \\
NGC 7331 ULX2 &             1.57 ($\pm$0.24)   & 1.27 ($\pm 0.22$)\\
NGC 7331 ULX3 & 2.03 ($\pm 0.03$)& 1.55 ($\pm 0.46$)\\ \hline
\end{tabular}
\end{table}

In the X-ray, two new sources were identified. The first is a bright, $10^{40}$ erg/s ULX in NGC~4157, associated with the central region of the galaxy, with a hard power-law spectrum. 
The second newly discovered source is a transient ULX in NGC~891 discovered in a 2ks observation with an observed X-ray luminosity of ~5$\times 10^{39}$ erg/s. Follow-up studies of archival XMM-Newton data (both prior to and after the \textit{Chandra} detection) place a consistent upper limit of ~2$\times 10^{38}$ erg/s.  
NGC 7331's ULXs have been previously studied by in X-ray and optical \citealt{2007Abolmasov}, and \citealt{2019Jin} performed a comprehensive study in X-ray of the total available archival data. Our results are in agreement with these studies, although we calculate the luminosity based on a smaller distance, which naturally returns fewer ULX candidates.

\section{Discussion}
\label{sec:conclusions}
We analyse a total of 9 \textit{Chandra} observations of seven edge-on spiral galaxies, identifying 12 ULX candidates, all of them associated with the host galaxy. 

\subsection{Comparison to Predicted ULX Populations}

The total average star formation rate (SFR) of these galaxies is 19.3 $M_{\odot}$/yr \cite{2021MNRAS.tmp.1552K}. With 12 ULX candidates identified, the ratio of ULXs per SFR is $\sim$ 0.62. This rate is close to both the expectation from the \citealt{2012MNRAS.419.2095M} luminosity function (0.60) as well as the scaling relation from \citealt{2020MNRAS.498.4790K}, (0.50). 

\subsection{NGC 891 ULX1 $-$ a bulge ULX?}
The brightest ULX in  NGC 891 is projected some distance ($\sim$13" or $\sim$ 600 pc) from the disk centre (Figure \ref{fig:im891}). This may indicate that the source formed in the bulge, rather than the disk. If so, a possible explanation, bolstered by the finding that the system shows evidence for a long duration transient \citep{2018ApJ...866..126H}, is that the system has an evolved donor with stable mass transfer, but an unstable accretion disk. Long period low mass X-ray binaries generally show long duration outbursts with peaks near the Eddington luminosity (e.g. \citealt{1998MNRAS.301..382S,2004MNRAS.355..413P}).  

\subsection{Transient ULXs}
The duty cycle and/or formation rate of transient ULXs are difficult to quantify because the discovery of transient ULXs is wholly serendipitous. However, studying the behaviour of transient ULXs may improve our knowledge of certain underlying physical mechanisms including the propeller effect, see, e.g., work by \citealt{2020arXiv200100642E,2020MNRAS.491.1260S}. 

This study identifies a transient ULX associated with the central region of NGC~891. Given the low numbers of transient ULXs that have been studied, and the scarcity of observations of NGC~891 ULX2 while it was active, understanding the physical mechanisms underlying the source is difficult. However, it merits comparison to the few better studied transient ULXs.

Given that it is located in the star-forming region of a spiral galaxy, it bears few similarities to the transient ULX source coincident with a globular cluster near NGC~1399 identified by \citealt{2010ApJ...721..323S}. The NGC~1399 source was active over a much longer baseline of at least 10 years, as it was observed in early ROSAT data (1993 and 1997) before turning off in 2003.  

Although the length of the observation in which NGC~891 ULX2 was active was very short ($\sim$ 2ks), it does rule out a similarity to the flaring ULXs detected by \citealt{2016Natur.538..356I}, as the lightcurve of this source shows no evidence of variability, and the observed X-ray luminosity is at least an order of magnitude lower than the two flaring ULXs. Similarly, the transient ULX source observed in M86 by \citealt{2013ApJ...779...14J} shows a peak luminosity of 6$\times 10^{42}$ erg/s, which is also orders of magnitude higher than the observed luminosity of NGC~891 ULX2. 

The clearest comparison NGC~891 ULX2 has to a transient ULX source is  NGC~300 ULX1, which was observed as a supernova imposter in 2010, and was consistently observed until the source fell below Swift's detection limit in 2019 \citep{2019MNRAS.488.5225V}. The source had a luminosity  of ~5$\times 10^{39}$ erg/s \citep{2018MNRAS.476L..45C} and exhibited pulsations \citep{2018ATel11179....1V}, and was observed to have a red supergiant donor \citep{2019ApJ...883L..34H}. The sources have similar measured peak luminosities of ~5$\times 10^{39}$ erg/s and may have been active for similar time scales. However, given the limited data available on the NGC~891 ULX2 source, it is difficult to draw further comparisons. 

While the sample of transient ULXs is small and difficult to study, the nature of transience in ULXs poses interesting questions for future studies, and in particular, quantifying the ULX population. Understanding the rich and diverse environments that form ULXs is the key to probe the cause and nature of transience in a subset of the sources.  

\subsection{Radial Distribution of X-ray Sources}
\begin{figure}
\includegraphics[width=8cm]{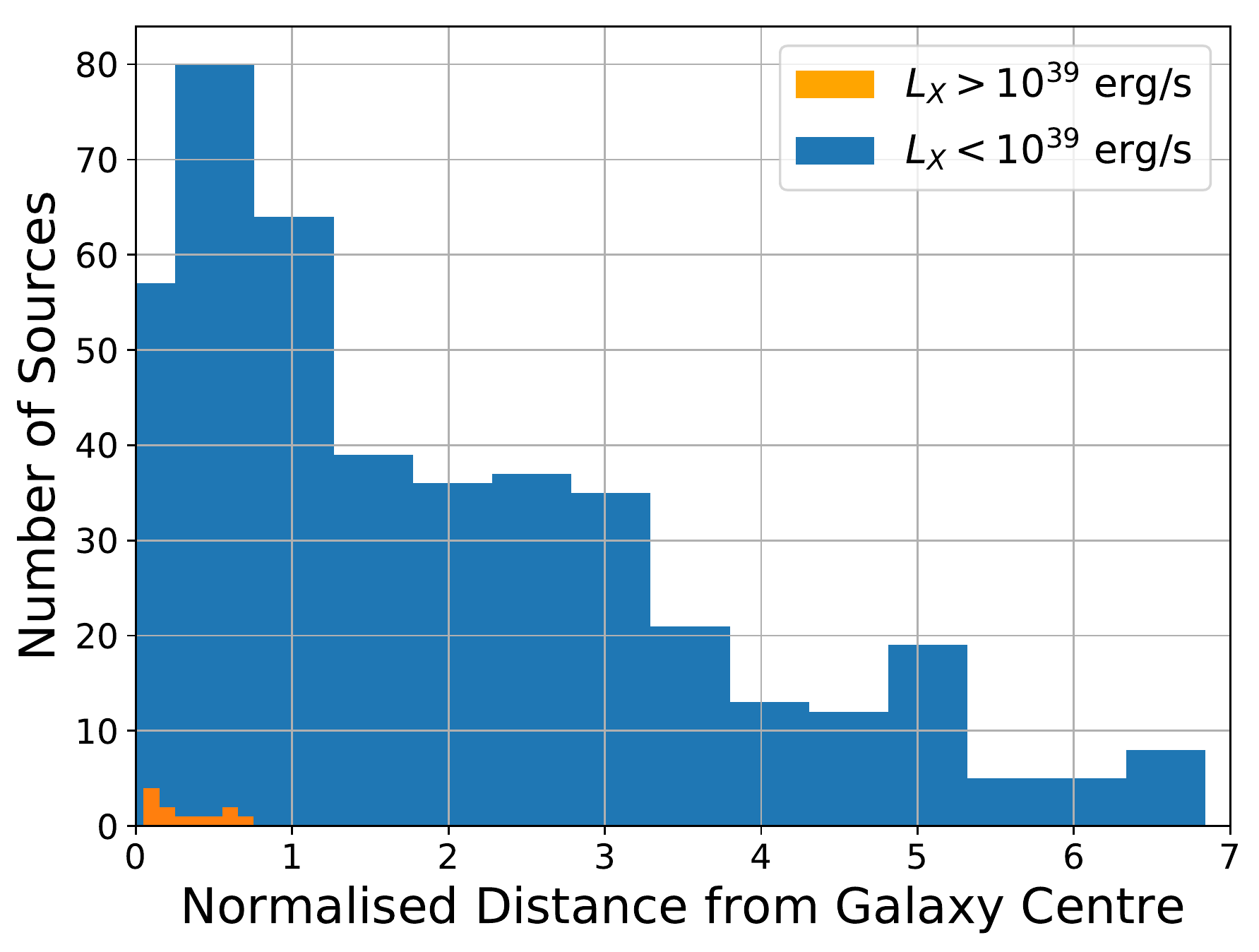}
\caption{X-ray point source (both those above $10^{39}$ erg/s and those below) distances from respective galaxy centres as measured by the 2MASS catalog \citealt{2006AJ....131.1163S} and scaled by $K_s$ isophotal diameter \citet{2003AJ....125..525J}. This, in combination with the optical images implies that the identified ULXs detected by this study are associated with the disk or bulge of their host galaxies.}
\label{fig:dists}
\end{figure}

 As seen in Figure \ref{fig:dists}, all of the identified ULXs are associated with the host galaxy, and within its isophotal radius. This figure also shows the radial distribution of the sources below the ULX limit (the Chandra exposures are sensitive down to a few $10^{37}$ erg/s, although we are less sensitive to off-axis sources). 
 
 All of the ULXs are associated with the central regions of the galaxy (within 1 isophotal radius). The non-ULXs also peak within the isophotal radius, but are distributed outside the galactic centre as well. Any contamination from background or foreground sources or more likely over the larger area outside of the central regions, so the true central concentration is at least as large as the strongly concentration population observed.
 
\subsection{Summary}
We present a study of 7 edge-on spiral galaxies and in which we identify 12 candidate ULXs among many detected X-ray point sources. Each ULX candidate is within one isophotal radius of the galaxy, and projected onto the central regions. This suggests that these ULXs are located in the disk or bulge, and not the halo. The ratio of ULXs per SFR for this sample is consistent with predictions of ULX populations, such as those made by  \citealt{2012MNRAS.419.2095M} and \citealt{2020MNRAS.498.4790K}. Lastly, we uncovered two transient ULXs, which may be long-duration transients. 

If long-duration transient ULXs are low mass X-ray binaries, as suggested by \citealt{1998MNRAS.301..382S,2004MNRAS.355..413P}, then one might expect to observe transient ULX sources in the halo. However, while the extant observations are less sensitive to detecting transient ULXs, the only two transient ULXs identified here are associated with the disk/bulge of their host galaxy, rather than the halo. Clearly a larger sample would aide our understanding of the population demographics.

\section*{Data Availability Statement}
The X-ray data in this article is available through the \textit{Chandra} archive\footnote{\url{https://cda.harvard.edu/chaser/}}, \textit{XMM-Newton} data archive \footnote{\url{https://www.cosmos.esa.int/web/xmm-newton/xsa}} and NuSTAR data archive \footnote{\url{https://heasarc.gsfc.nasa.gov/docs/nustar/nustar_archive.html}}. The optical data is available through the ESO Online Digitized Sky Survey \footnote{\url{https://archive.eso.org/dss/dss}}. Archival HST data can be found at the Hubble Legacy Archive, hosted by the Space Telescope Science Institute (STScI), the Space Telescope European Coordinating Facility (ST-ECF), and the Canadian Astronomy Data Centre (CADC).\footnote{\url{http://hla.stsci.edu/}}.

\section*{Acknowledgements}


We thank our referee, Andreas Zezas, for a helpful report which improved the quality and presentation of the results in this paper. KCD and DH acknowledge funding from the Natural Sciences and Engineering Research Council of Canada (NSERC), the Canada Research Chairs (CRC) program, and the McGill Bob Wares Science Innovation Prospectors Fund. KCD acknowledges fellowship funding from the McGill Space Institute. KK acknowledges support from the the Swiss National Science Foundation Professorship grant (project number PP00P2 176868). AK acknowledges support from NASA through grant number GO-14738 from STSci and Chandra award GO-11111. JS and NV acknowledge support from the Packard Foundation. RU acknowledges support from Chandra award GO7-18032A and HST award HST-GO14351. SEZ acknowledges support from  JPL RSA No. 1659290. KCD thanks Breanna Binder, McKinley Brumback, Francesa Fornasini, and Rich Plotkin for helpful discussion.  

The scientific results reported in this article are based
on observations made by the Chandra X-ray Observatory,
NuSTAR Observatory, the XMM-Newton Observatory, and the ESO Online Digitized Sky Survey.
The following software and packages were used for analysis: \textsc{ciao}, software provided by the Chandra X-ray Center (CXC),   \textsc{heasoft} obtained from the High Energy Astrophysics Science Archive Research Center (HEASARC), a service of the Astrophysics Science Division at NASA/GSFC and of the Smithsonian Astrophysical Observatory's High Energy Astrophysics Division, SAOImage DS9, developed by Smithsonian Astrophysical Observatory, \textsc{NuSTARDAS}  jointly developed by the ASI Science Data Center (ASDC, Italy) and the California Institute of Technology (Caltech, USA),  \textsc{numpy} \citep{2011arXiv1102.1523V},  \textsc{matplotlib} \citep{2007CSE.....9...90H} and \textsc{MaLTPyNT} \citep{2015ascl.soft02021B}. This publication makes use of data products from the Two Micron All Sky Survey, which is a joint project of the University of Massachusetts and the Infrared Processing and Analysis Center/California Institute of Technology, funded by the National Aeronautics and Space Administration and the National Science Foundation. Part of this analysis is based on observations obtained with XMM-Newton, an ESA science mission with instruments and
contributions directly funded by ESA Member States and
NASA. We  also acknowledge use of NASA's Astrophysics Data System and Arxiv.








\bibliographystyle{mnras}

\bibliography{eos}
\bsp	
\label{lastpage}
\end{document}